\documentclass[a4paper,
				keeplastbox,   
			   hyphens,      
              ]{jacow}

\makeatletter%
	\ifboolexpr{bool{xetex}}
	 {\renewcommand{\Gin@extensions}{.pdf,%
	                    .png,.jpg,.bmp,.pict,.tif,.psd,.mac,.sga,.tga,.gif,%
	                    .eps,.ps,%
	                    }}{}
\makeatother

 {\usepackage[utf8]{inputenc}}           

\ifboolexpr{bool{xetex} or bool{luatex}} 
 {}                                      

\usepackage[USenglish]{babel}			 
\ifboolexpr{bool{jacowbiblatex}}%
 {%
  \addbibresource{jacow-test.bib}
  \addbibresource{biblatex-examples.bib}
 }{}
\begin{document}

\title {Compact ring-based X-ray source with
		on-orbit and on-energy laser-plasma
		injection}

\author{Marlene Turner\textsuperscript{1}\thanks{marlene.turner@cern.ch}, CERN, Geneva, Switzerland\\
		Jeremy Cheatam, Auralee Edelen, CSU, Fort Collins, Colorado, USA \\
		Osip Lishilin, DESY Zeuthen, Zeuthen, Germany\\
		Aakash Ajit Sahai, Imperial College Physics, London, Great Britain\\
		Andrei Seryi, JAI, Oxford, Great Britain\\
		Brandon Zerbe, MSU, East Lansing, MI, USA \\
		Andrew Lajoie, Chun Yan Jonathan Wong, NSCL, East Lansing, MI, USA \\
		Kai Shih, SBU, Stony Brook, New York\\
		James Gerity, Texas A\&M University, College Station, TX, USA\\
		Gerard Lawler, UCLA, Los Angeles, CA, USA \\
		Kookjin Moon, UNIST, Ulsan, Korea\\
		\textsuperscript{1}also at Technical University of Graz, Graz, Austria}
	
\maketitle

\begin{abstract}
We report here the results of a one week long investigation into the conceptual design of an X-ray source based on a compact ring with on-orbit and on-energy laser-plasma accelerator (mini-project 10.4 from \cite{UNIFYINGPHYSICS}). We performed these studies during the June 2016 USPAS class "Physics of Accelerators, Lasers, and Plasma\ldots" applying the art of inventiveness TRIZ \cite{TRIZ}. We describe three versions of the light source with the constraints of the electron beam with energy $1\,\rm{GeV}$ or $3\,\rm{GeV}$ and a magnetic lattice design being normal conducting (only for the $1\,\rm{GeV}$ beam) or superconducting (for either beam). The electron beam recirculates in the ring, to increase the effective photon flux.  We describe the design choices, present relevant parameters, and describe insights into such machines.
\end{abstract}

\section{INTRODUCTION}
Laser wakefield acceleration \cite{Laser-plasma-theory} experiments (LWFA) achieved GeV electron energies in cm-scales using plasma waves \cite{Pukhov-laser-bubble, cavitation-laser-expt-1, cavitation-laser-expt-2}. In this paper we explore a compact synchrotron light source, that uses a laser wakefield accelerated electron beam in combination with a conventional lattice. We outline a design that produces \SI{0.4}{keV} (water-window) or \SI{10}{keV} photons and we estimate the design parameters of the compact light source, the achievable brilliance, and we discuss the feasibility and challenges of the design. 
Our design uses the state of the art technology including a laser plasma gas-jet accelerator, a quadrupole doublet to focus and confine the beam, four 90 degree dipole bending magnets (super-conducting or normal conducting) to keep the beam on a periodic lattice, and a \SI{2}{m} long wiggler magnet to produce the desired radiation. We do not study the radiation emitted by the bending magnets, or the betatron oscillations of the electron beam in the plasma bubble. 

A schematic of the design is in Fig. \ref{schematics}, and the investigated design criteria are detailed below: 
\begin{Itemize}
\item \SI{0.4}{keV} photons produced by a \SI{1}{GeV} electron beam and normal-conducting  magnets.
\item \SI{0.4}{keV} photons produced by a \SI{1}{GeV} electron beam and super-conducting  magnets.
\item \SI{10}{keV} photons produced by a \SI{3}{GeV} electron beam and super-conducting  magnets.
\end{Itemize}

\section{DESIGN OF THE MACHINE}
\subsection{Plasma Based Self-injected Electron Accelerator}
\begin{figure}
 \includegraphics[width = \columnwidth]{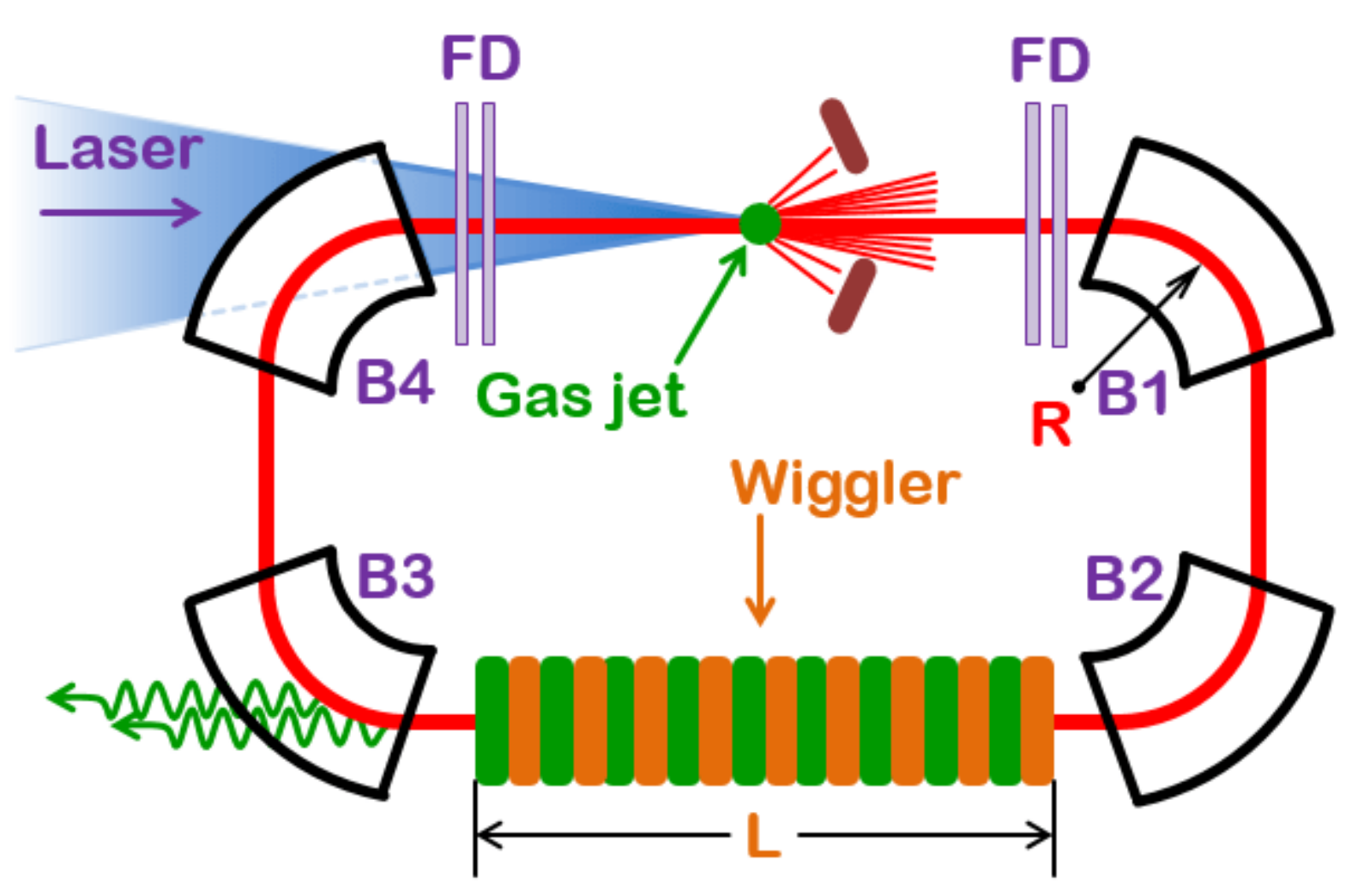}
\caption{Schematics of the compact ring design. The laser beam enters through a window in the B4 dipole bending magnet, ionizes the gas jet and creates strong plasma wakefields to self-inject and accelerate electrons. Furthermore the produced electron beam gets parallelized with a quadrupole focusing doublet (FD). The electron beam is held on a circular trajectory by four 90-degree bending magnets (B1-B4). Opposite to the plasma injector, a Wiggler magnet produces the desired radiation. A second FD is used to refocus the circulated beam.}
	\label{schematics}
\end{figure}
The laser wakefield acceleration with self-guiding and self-injection provides the electron beam for the compact ring. An intense \SI{500}{TW}-class Ti:Sapphire CPA system is injected through a window in one of the bending dipoles (B4 in Fig. \ref{schematics}), impinges on a gas jet, ionizes the gas and creates strong plasma wakefields that accelerate the self-injected electron beam to GeV-energies. The chosen laser and plasma parameters are summarized in Tables \ref{laserparameters} and \ref{plasmaparameters}.
Based on existing systems, we choose laser parameters achievable in the near future \cite{Laser-system}. A Ti:Sapphire laser system with a laser wavelength of $\lambda_l = 780\,\rm{nm}$, a pulse duration of $45\,\rm{fs}$, a maximum laser power of $400\,\rm{TW}$, $30\,\rm{J}$ of energy per pulse, and a laser spot size radius of 37\,{\textmu}m. The laser intensity $I$ is then approximately $10^{19}\,\rm{W/cm}^2$ which corresponds to a laser strength parameter $a_0$ of about $2.1$ for a Gaussian radial laser distribution. 
\begin{equation}
a_0 \approx \left( \frac{I [\rm{W/cm}^2]}{1.37 \times 10^{18}} \right) ^{\frac{1}{2}} \cdot \lambda_l \ [\text{\textmu m}]
\end{equation}
The chosen plasma density of the gas-jet is $n=1.75 \times 10^{17}\,\rm{cm}^{-3}$ and optimizes the achievable maximum electron energy. The laser frequency $\omega_l = 2.4\times 10^{15}\,\rm{rad/s}$ is above the critical plasma frequency $\omega_{pc} = 2.5\times 10^{13}\,\rm{rad/s}$.

The depletion $L_{dpl}$ and dephasing $L_{dph}$ length \cite{REVIEW} for these laser-plasma parameters are estimated with: 
\begin{equation}
L_{dpl} = \frac{1}{2 a_0} \frac{\lambda_p^3}{\lambda_l^2}\, , 
L_{dph} = \frac{1}{4} \frac{\lambda_p^3}{\lambda_l^2}
\end{equation}
to be $L_{dpl}\approx 17\,\rm{cm}$ and $L_{dph}\approx 20\,\rm{cm}$, thus $17\,\rm{cm}$ is the upper bound on our total possible acceleration length. The maximum accelerating gradient $E_{max}$ is estimated based on the cold plasma wave-breaking field (\ref{eq:wavebreaking}) to be \SI{400}{MeV/cm}.
\begin{equation}
\label{eq:wavebreaking}
E_{max} \sim  \frac{m c \omega_p}{e}
\end{equation}
This means that our target energies of $1\,\rm{GeV}$ and $3\,\rm{GeV}$ can be reached with an acceleration length of $2.5\,\rm{cm}$ ($\sim 4.4$ Rayleigh lengths $z_r$) and $7.5\,\rm{cm}$ ($\sim 13.1\,z_r$). This acceleration length is longer than gas jets in use presently ($\approx$ \SI{5}{mm}) and may require a novel gas jet setup, particularly for the $3\,\rm{GeV}$ beam.

In this design, we rely on analytic estimates rather than on the results of 3D simulations of the setup. Consequently, for the remainder of the ring calculations we used electron beam parameters that are typically achievable with similar laser wakefield acceleration stages \cite{LEEMANS}. The electron beam size is chosen to be $\sigma_r \approx \sigma_z \approx 1 \frac{c}{\omega_p} \approx $12\,{\textmu}m. The electron energy spread is $\frac{\Delta E}{E_0} = 2\%$. The beam divergence is chosen to be $\sigma_\theta = 0.5\,\rm{mrad}$. Reasonable bunch charge for the \SI{1}{GeV} and \SI{3}{GeV} electron beams are \SI{10}{pC} and \SI{7}{pC}\cite{10PC}.
\begin{table}[hbt]
   \centering
   \caption{Ti:sapphire Laser Parameters}
   \begin{tabular}{lc}
       \toprule
Laser wavelength & \SI{780}{nm}\\
Laser power & \SI{379}{TW}\\
Spot size radius &  \SI{37}{\micro m} \\
Intensity & $10^{19}\,\rm{W/cm}^2$ \\
$a_0$ & \SI{2.1}{} \\
Laser pulse length (FWHM) & \SI{45}{fs} \\
Pulse reprate ($f_{\rm{rep}}$) &  	\SI{1}{Hz} \\
Pulse Energy & \SI{30}{J} \\
       \bottomrule
   \end{tabular}
   \label{laserparameters}
\end{table}

\begin{table}[hbt]
   \centering
   \caption{Plasma Parameters}
   \begin{tabular}{lc}
       \toprule
Plasma density & \SI{1.75e17 }{cm^{-3}}\\
Accelerating gradient & \SI{0.4}{GeV/cm} \\
Bubble radius &  \SI{37}{\micro m} \\
Depletion length & \SI{16.9}{cm} \\
Dephasing length & \SI{20}{cm} \\
Acceleration length for 1 GeV & \SI{2.4}{cm} \\
Acceleration length for 3 GeV & \SI{7.2}{cm} \\
       \bottomrule
   \end{tabular}
   \label{plasmaparameters}
\end{table}

\subsection{Magnet Design}
The design of the compact ring (see Figure \ref{schematics}) is laid out as four $90\,\rm{-degree}$ sector dipoles, either \SI{10}{T} super-conducting (s.c.) or \SI{1.5}{T} normal conducting (n.c.). Parameters for the bending magnets are shown in Table \ref{magnets}. The laser-plasma accelerator system is located in the middle of a \SI{2}{m} drift, enclosed by a pair of focusing quadrupole doublets (strengths $\sim$ \SI{50}{T/m}, \SI{100}{T/m}) used to begin transport of the diverging electron beam. 
The focusing strengths of the quadrupole magnets were estimated to parallelize/focus an electron beam with parameters outlined in the previous section. The doublet upstream the gas-jet focuses the electron beam to avoid interaction with residual plasma (see Discussion section). Dipole magnets provide weak- and edge- focusing in the horizontal plane. Assuming the betatron tune to be $Q = 0.3$, we can estimate the average beta function $\overline{\beta}$ using:
\begin{equation}
2\,\pi\,Q = \oint \frac{ds}{\beta(s)} = \frac{C}{\overline{\beta}}
\end{equation}
where $C$ is the circumference of the ring. The horizontal $\Delta x$ and vertical $\Delta y$ beam sizes are estimated based on the bending radius $\rho$ and the circumference $C$. The average dispersion function $\overline{D}$ was evaluated based on Eq.~(\ref{eq:disp}):
\begin{equation}
\label{eq:disp}
\overline{D} \approx \rho
\end{equation}
We performed a MAD-X \cite{MADX} optics calculation after the school to confirm these estimates. 

The wiggler is located in the drift space between B2 and B3, and the remaining two drift spaces are each \SI{0.5}{m} leaving enough space for limited diagnostics and collimators. The electron energy loss per turn is dominated by the synchrotron radiation loss $E_{sr}$ in the bending magnets and ranges from \SI{40}{keV} to \SI{7}{MeV}. The energy radiated in the Wiggler magnet $E_{\rm{rad}}$ is listed in Table  \ref{tab:radiation}.
There is no re-accelerating section in our compact ring design. This means that the beam loses energy and drifts off from the central orbit. Therefore, the number of turns $N_{\rm{turns}}$ is limited by the selected \SI{10}{cm} horizontal apertures $x_{\rm{max}}$ of the bending magnets.
The number of turns was calculated according to Eq.~(\ref{eq:Nturns}), by taking the $2\sigma_r$ radial electron beam size, and calculating how much energy can be lost before the beam touches the horizontal magnet aperture.
\begin{equation}
\label{eq:Nturns}
N_{turns} =  \frac{(x_{max} - 2 \Delta x)}{\overline{D}} \frac{E_{\rm{beam}}}{E_{sr}\rm{/turn}}
\end{equation}

\begin{table}[hbt]
   \centering
   \caption{Optics- and Dipole Magnet Parameters}
   \setlength\tabcolsep{4pt}
   \begin{tabular}{lccc}
       \toprule
\textbf{Parameter} & \textbf{\SI{1}{GeV} n.c.} & \textbf{\SI{1}{GeV} s.c.} & \textbf{\SI{3}{GeV} s.c.}\\
Ben. field  & 1.5 T & 10 T & 10 T\\
Ben. radius $\rho$ & 2.23 m & 0.33 m & 1.02 m\\
C & 19.1 m & 7.1 m & 9.4 m\\
$\overline{\beta}$ & 10.1 m & 3.8 m & 5 m\\
$\Delta x$/$\Delta y$ & 4.5/0.05 cm & 0.68/0.05 cm & 2/0.05 cm\\
$E_{sr}$/turn & 40 keV & 260 keV & 7 MeV\\
$N_{\rm{turns}}$ & 62 & 490 & 13\\
$\overline{D}$ & 2.23 m& 0.33 m& 1.02 m\\
       \bottomrule
   \end{tabular}
   \label{magnets}
\end{table}

\begin{table}[tb!]
  \centering
  \setlength\tabcolsep{4pt}
  \caption{Chosen (upper section) and Derived (lower section) X-ray Source Parameters}
  \label{tab:radiation}
  \begin{tabular}{lccc}
    \toprule
		\textbf{Parameter} & \textbf{\SI{1}{GeV} n.c.} & \textbf{\SI{1}{GeV} s.c.} & \textbf{\SI{3}{GeV} s.c.}\\
    $\text{E}_\text{ph}$ (\text{keV}) & $0.4$ & $0.4$ & $10.0$ \\
    $\lambda_\text{wiggler}$ (\text{mm}) & $15$ & $15$ & $100$ \\
    \midrule
    $\text{B}_\text{wiggler} (\text{T})$ \cite{SupercondWiggler, Wiggler}& $0.60$ & $0.60$ & $1.7$ \\
    $\text{K}$ & $0.84$ & $0.84$ & $16$ \\
    $\text{E}_\text{rad} (\text{keV})$ & $2.1$ & $2.1$ & $140$ \\
    $B_{\rm{e-life}} \left ( \frac{\text{photons}}{\text{mm}^2 \text{mrad}^2 \text{s}} \right )$ & $2 \times 10^{10}$ & $9  \times 10^{10}$ & $1  \times 10^{10}$ \\
    $B_{\rm{spill}} \left ( \frac{\text{photons}}{\text{mm}^2 \text{mrad}^2 \text{s}} \right )$& $5 \times 10^{15}$ &  $8 \times 10^{15}$ & $3  \times 10^{15}$ \\
    $\text{train durat.} ( \text{\textmu s})$ & $3.9$ & $12$ & $0.47$\\
    \bottomrule
  \end{tabular}
\end{table}

\subsection{Radiation Production}
\label{radiation}
Table \ref{tab:radiation} details the radiation source parameters for all considered design versions.  We choose a length of $L = 2\,\rm{m}$ for the undulating magnetic field region, where the electron bunch emits an approximately \SI{6.7}{ns}-long X-ray pulse at each turn through the B3 aperture before beam loss. As shown in the table, we designed the X-ray energy $\text{E}_\text{ph}$ to either be within the water window ($0.4\,\text{keV}$) or to be fairly hard ($10\,\text{keV}$) depending on the energy of the injected electron beam.  The chosen parameters result in the source being near the wiggler/undulator transition or well within the wiggler regime, respectively, as can be seen by the parameter K. The brilliance per electron passage in the wiggler $B_{\rm{spill}}$ was estimated based on Eq.~(\ref{eq:brilliance}):
\begin{equation}
\label{eq:brilliance}
B_{\rm{spill}} = \frac{\rm{N}_{\rm{ph}} L}{\pi \Delta x \Delta y \theta_{\rm{v} } \theta_{\rm{h}} \Delta t} \cdot \rm{N}_{\rm{elec} }
\end{equation}
and the brilliance per electron lifetime $B_{\rm{e-life}}$:
\begin{equation}
\label{eq:elife}
B_{\rm{e-life}} = \frac{\rm{N}_{\rm{ph}} L}{\pi \Delta x \Delta y \theta_{\rm{v}} \theta_{\rm{h}}} \cdot \rm{N}_{\rm{elec}} \cdot f_{\rm{rep}}\cdot \rm{N}_{\rm{turns}}
\end{equation}
where $\rm{N}_{\rm{ph}}$ is the number of photons emitted per unit length (from Eq.~(3.22) of \cite{UNIFYINGPHYSICS}), $\text{N}_{\text{elec}}$ is the number of electrons in the beam, $f_{\rm{rep}}$ is the frequency of electron injection (\SI{1}{Hz}) and $\Delta t$ the time per revolution. The emission angles $\theta_{\rm{h}}$ and $\theta_{\rm{v}}$ are estimated with $\theta_{\rm{h}} = \Delta x/(2C)$ and $\theta_{\rm{v}} = \Delta y/(2C)$.

Of the chosen parameter sets, the \SI{1}{GeV} electrons with the superconducting bending magnets (\SI{10}{T}) resulted in the largest brightness. We also want to note that these brightness estimates include only the radiation from the B3 aperture, not including the synchrotron radiation produced by B3.

\section{DISCUSSION}

One of the concerns of the re-circulating design is that the gas/plasma from the injector resides in the lattice longer than it takes the electron beam to circulate in the ring. The gas flow velocity from typical gas jets is $M = 20\,\rm{km/s}$ \cite{GASFLOW}, and one electron beam turn in a \SI{10}{m} circumference ring takes $\sim$ \SI{30}{ns}.  Thus the gas flow advances by $\sim$ \SI{600}{\micro m} and will therefore still be in the path of the beam for much of its residence time. Taking $\sim 10^{14}\,\rm{W/cm}^2$ as ionization threshold of the plasma and assuming a radial laser size of \SI{35}{\micro m} gives a plasma channel radius of \SI{140} {\micro m}. This estimate suggests that the plasma column has already propagated far enough to eliminate the possibility of any further interaction with the beam. So, the electron beam is only likely to interact with the residual gas and not the laser-ionized plasma column. Thus, the interaction process of the beam is dominated by random scattering off the gas, which leads to a divergence growth of the order of micro-rad, as opposed to a coherent lensing of the order of milli-rad by the plasma \cite{beam-plasma-interaction}. Electron beam of pC charge with a beam size of $\sigma_r = 5$\,{\textmu}m and $\sigma_z = 10$\,{\textmu}m is not dense enough to ionizes the gas.

Apart from using the radiation created by the wiggler, it would also be possible to use the radiation from the bending magnets or the X-rays produced by the betatron oscillations of the electron beam in the bubble \cite{laser-plasma-betatron}. A Compton light source could also be realized by making the electron bunch collide with a laser pulse.

\section{CONCLUSIONS}
In this paper we proposed different design solutions for a compact ring-based X-ray source with an on-orbit and on-energy laser-plasma accelerator. We considered four 90 degrees bending magnets to keep the particles on a circular orbit and the desired radiation is created by a wiggler magnet. The peak brilliance is as high as $8 \times10^{15}\,\rm{photons/(mm}^2 \rm{mrad}^2 \rm{s})$. Even though modern, 3rd and 4th generation, light sources routinely create a peak brilliance in the order of $10^{25}\,\rm{photons/(mm}^2 \rm{mrad}^2 \rm{s}\,0.1\%\rm{BW})$ \cite{DESY}, this design can be considered attractive due to its compactness and small footprint. This design would be suitable for use in, for example, a university setting.
Future work should include plasma simulations to obtain better estimates for the electron beam parameters, optics optimization and particle tracking to obtain electron beam sizes, studies on the gas-jet design, further investigations on the electron scattering on the residual gas, and a more detailed description of the magnet design.


\begin{thebibliography}{99}
\newcommand{\href}[2]{#2, \url{#1}}
\renewcommand{\textit}[1]{\enquote{#1}}

\bibitem{UNIFYINGPHYSICS} A. Seryi, 
	\textit{Unifying physics of accelerators, lasers and plasma}, 
	\href{https://www.crcpress.com/Unifying-Physics-of-Accelerators-Lasers-and-Plasma/Seryi/p/book/9781482240580}{CRC Press, 2015}.
	
\bibitem{TRIZ}
	E. Seraia and A. Seryi,
\textit{Accelerating Science TRIZ inventive methodology in illustrations},
\href{http://arxiv.org/abs/1608.00536}{arXiv:1608.00536}.

\bibitem{Laser-plasma-theory} 
	T. Tajima and J. M. Dawson,
	\textit{Laser Electron Accelerator}, 
	\href{http://link.aps.org/doi/10.1103/PhysRevLett.43.267}{Phys. Rev. Lett. \textbf{43}, pp.267-270, 1979}.
	
\bibitem{Pukhov-laser-bubble}
	A. Pukhov and J. Meyer-Ter-Vehn,
	\textit{Laser wake field acceleration: the highly non-linear broken-wave regime},
	\href{http://dx.doi.org/10.1007/s003400200795}{Appl. Phys. B \textbf{74}, pp.355-361, 2002}. 
	
\bibitem{cavitation-laser-expt-1}
	S. P. D. Mangles \emph{et al.},
	\textit{Monoenergetic beams of relativistic electrons from intense laser-plasma interactions},
	\href{http://dx.doi.org/10.1038/nature02939}{Nature \textbf{431}, 535, 2004}
	
\bibitem{cavitation-laser-expt-2}
	C. G. R. Geddes \emph{et al.},
	\textit{High-quality electron beams from a laser wakefield accelerator using plasma-channel guiding},
	\href{http://dx.doi.org/10.1038/nature02900}{Nature \textbf{431}, 538, 2004}

\bibitem{Laser-system} 
	V. Yanovsky \emph{et al.},
	\textit{Ultra-high intensity 300-TW laser at 0.1 Hz repetition rate}, 
	\href{http://dx.doi.org/10.1364/OE.16.002109}{Vol. 16, Issue 3, pp. 2109-2114, 2008},  

\bibitem{REVIEW}
	E. Esarey \emph{et al.},
	\textit{Physics of laser-driven plasma-based electron accelerators},
	\href{http://dx.doi.org/10.1103/RevModPhys.81.1229}{Rev. Mod. Phys. 81, 1229}.


\bibitem{LEEMANS}
	W. P. Leemans \emph{et al.},
	\textit{Multi-GeV Electron Beams from Capillary-Discharge-Guided Subpetawatt
	Laser Pulses in the Self-Trapping Regime},
	\href{http://dx.doi.org/10.1103/PhysRevLett.113.245002}{Phys. Rev. Lett. 113, 245002}.

\bibitem{10PC}
	S. Kneip \emph{et al.},
	\textit{Near-GeV Acceleration of Electrons by a Nonlinear Plasma Wave Driven
	by a Self-Guided Laser Pulse},
	\href{http://journals.aps.org/prl/abstract/10.1103/PhysRevLett.103.035002}{Phys. Rev. Lett. 103, 035002}.

\bibitem{MADX}
	H. Grote and F. Schmidt,
	\textit{MAD-X -- An Upgrade from MAD8},
	CERN-AB-2003-024  ABP.

\bibitem{SupercondWiggler}
	N. A. Mezentsev,
	\textit{Survey of superconducting insertion devices for light sources},
	\href{http://accelconf.web.cern.ch/AccelConf/p05/PAPERS/TOAA003.PDF}{Proceedings of 2005 Particle Accelerator Conference, Knoxville, Tennessee}.

\bibitem{Wiggler}
	Y. Ivanyushenkov \emph{et al.},
	\textit{Development and operating experience of a short-period superconducting undulator at the Advanced Photon Source},
	\href{http://dx.doi.org/10.1103/PhysRevSTAB.18.040703}{Phys. Rev. ST Accel. Beams 18, 040703}.

\bibitem{GASFLOW} 
	S. Semushin and V. Malka,
	\textit{High density gas jet nozzle design for laser target production}, 
	\href{http://dx.doi.org/10.1063/1.1380393}{Review of Scientific Instruments 72, 2961, 2001} 

\bibitem{beam-plasma-interaction}
	A. A. Sahai,
	\textit{Beam interaction with Plasma-Vacuum interface}, 
	\href{http://indico.cern.ch/event/269506/contributions/607723/}{CLIC Experimental \& Breakdown Studies Meeting, August 2013.}

\bibitem{laser-plasma-betatron} 
	A. Rousse \emph{et al.},
	\textit{Production of a keV X-ray beam from synchrotron radiation in relativistic laser-plasma interaction},
	\href{http://dx.doi.org/10.1103/PhysRevLett.93.135005}{Phys. Rev. Lett. 93, 135005,  2004}.
	
\bibitem{DESY}
	M. Abd-Elmeguid \emph{et al.},
	\textit{TESLA, Technical Design Report}.

\end{thebibliography}
\end{document}